# Metastable states in plateaus and multi-wave epidemic dynamics of Covid-19 spreading in Italy

Gaetano Campi[1,2], Maria Vittoria Mazziotti[2], Antonio Valletta[3], Giampietro Ravagnan[4], Augusto Marcelli[5], Andrea Perali[6], Antonio Bianconi[1,2,7]

[1] Institute of Crystallography, Consiglio Nazionale delle Ricerche CNR, via Salaria Km 29.300, Monterotondo Roma, I-00015, Italy;
[2] Rome International Centre Materials Science Superstripes RICMASS via dei Sabelli 119A, 00185 Rome, Italy;
[3] Institute for Microelectronics and Microsystems IMM, Consiglio Nazionale delle Ricerche CNR Via del Fosso del Cavaliere 100, 00133 Roma, Italy;
[4] Istituto di Farmacologia Traslazionale IFT, Consiglio Nazionale delle Ricerche CNR Via del Fosso del Cavaliere 100, 00133 Roma, Italy;
[5] INFN - Laboratori Nazionali di Frascati, 00044 Frascati (RM) Italy;
[6] School of Pharmacy, Physics Unit, University of Camerino, 62032 Camerino (MC) Italy.
[7] National Research Nuclear University MEPhI (Moscow Engineering Physics Institute), 115409 Moscow, Russia

**Abstract**

The control of Covid 19 epidemics by public health policy in Italy during the first and the second epidemic waves has been driven by using reproductive number $R_t(t)$ to identify the *supercritical* (percolative), the *subcritical* (arrested), separated by the *critical* regime. Here we show that to quantify the Covid-19 spreading rate with containment measures (CSRwCM) there is a need of a 3D expanded parameter space phase diagram built by the combination of $R_t(t)$ and doubling time $T_d(t)$. In this space we identify the dynamics of the Covid-19 dynamics Italy and its administrative Regions. The supercritical regime is mathematically characterized by *i)* the power law of $T_d$ vs. $[R_t(t)-1]$ and *ii)* the exponential behaviour of $T_d$ vs. time, either in the first and in the second wave. The novel 3D phase diagram shows clearly *metastable states* appearing before and after the second wave critical regime. for loosening quarantine and tracing of actives cases. The metastable states are precursors of the abrupt onset of a next nascent wave supercritical regime. This dynamic description allows epidemics predictions needed by policymakers to activate non-pharmaceutical interventions (NPIs), a key issue for avoiding economical losses, reduce fatalities and avoid new virus variant during vaccination campaign.

## 1. Introduction

While the laws of uncontrolled epidemics spreading in a single network are well known[1] the quantitative description of the epidemics dynamics in multilayer heterogeneous networks[2] with containment measures is a strategic hot topic for statistical physics of living matter[3-10] to face Covid-19 pandemia showing non-uniform space population density and short time ~~time~~ heterogeneity,[11-15] which give a epidemic dynamics characterized by multiple waves where supercritical phases are intercalated by metastable states in plateaus due to intermittent weakening of the lockdown, quarantine and tracing rules in country enforcing the Lockdown Stop and Go (LSG) policy.[16-19]

In the second year of Covid-19 epidemic spreading it is mandatory both to avoid the onset of the third wave and to support an efficient vaccine immunization strategy. The non-medical containment measures need to be addressed to reduce the number of infected cases to minimize the probability of lethal virus mutations in the huge number of infected cell fission processes and to reduce the fatalities number, before the immunization with vaccines is obtained. In Italy, at the end of the second 2020 Covid epidemic wave while several researches have analysed short-time intervals of the epidemic spreading,[19-24] there is a lack of information on the dynamics of the full-time window of the first and the second Covid-19 waves. In the time evolution of the Covid-19 spreading rate with containment measures (CSRwCM) three main regimes have been clearly identified in ref 15: the *supercritical, critical,* and the *subcritical regime.* In the *supercritical* phase the extrinsic effects control the characteristic time s in the exponential law of the time-dependent doubling time

$$T_d(t)=Ae^{(t-t_0)/s} \qquad (1)$$

moreover, we have verified ~~here~~ that this phase is characterized by the power law function

$$T_d(t)=C(R_t(t)-1)^{-v} \qquad (2)$$

of the variable doubling time $T_d(t)$ vs. the reproductive number $R_t(t)$. In this supercritical regime, the cumulative curve of the total number of cases of the epidemics, approaches the critical regime, following the complex Ostwald growth law,[23,24] which is a mixed exponential and power-law behavior determined by nucleation and growth of different phases in out of equilibrium complex multiphase systems.[25,26]

We have verified the physical laws of the time evolution of the CEwCM using the new 3D expanded parameter space $T_d(t, R_t)$ to describe the time evolution of the two Covid-19 epidemic waves, mandatory to face the onset of the third wave. In addition, we introduce here the new parameter RTD that takes into account both $R_t(t)$ and $T_d(t)$. In this 3D expanded parameter space, arrested metastable phases, with $R_t>1$ and $T_d(t)>40$ days in the subcritical and critical regime, are precursors of the possible onset of the supercritical regime of the third wave. In this work we identify and show the metastable states precursors observed in Italy at the end of the second wave in 2021. Nowadays, these states are well identified and we expect that the situation may rapidly evolve, depending on the enforced containment policy rules, toward either a third wave in the supercritical regime or in the arrested subcritical regime. The results provide a quantitative evaluation of the Covid-19 evolution trough different phases resulting from different containment policies. The method gives us the possibility to foresee the evolution of the pandemic, and through the identification of metastable states in the critical regime, help policymakers to avoid the occurrence of new pandemic waves.

## 2. Results and Discussion

Data for each country have been taken from the recognized public database *OurWorldInData*[27]. We have initially extracted the time-dependent doubling time $T_d(t)$ from the curve of total infected cases, $Z(t)$, and, after, the time-dependent reproductive number $R_t(t)$ from the curve of active infected cases, $X(t)$, using the methodological definition provided by the Koch Institute[28] in Germany, as described in ref. 15. In a previous work[15] we have verified the results of this approach by using the inverted SIR theory where the effective reproductive number $R_e(t)$ and $T_d(t)$ have been extracted from joint $Z(t)$ and $X(t)$ curves.

*2.1 Metastable phases and dynamics of the second wave in Italy and Germany*
Figure 1 shows the 3D phase diagram ($T_d$, $R_t$, t) of the epidemic spreading in Italy and Germany from 1st January 2020. The grey strip corresponds to the *critical* regime occurring above the *supercritical* area for $T_d(t)$ between 40 and 100 days. Here we can easily visualize the supercritical exponential growth, described by (1), of both the 1st and 2nd wave [Fig. 1a and Fig. 1b] that follow the linear behavior in the semi-logarithmic plot of $T_d$ versus $R_t$ and $R_t$-1 [see dashed arrows]. The rate of the growth described by the *s*-factor is different due to the different policy containment measure enforced during the two waves. Above the critical region, $T_d$ becomes large enough ($T_d$>100) and the exponential growth is arrested. In this arrested *subcritical* phase $R_t$ becomes < 1, as can be seen in the semi-logarithmic projection ($T_d$, $R_t$-1) in Fig. 1b where values $R_t$<1 are not showed. After the arrested phase, if $T_d$ decreases and $R_t$ increases, new states appear in the region with $R_t$ >1 and $T_d$ around 100 over the critical phase. They give rise to a phase, metastable, intermediate between the arrested phase and the occurrence of a new pandemic wave. In this transition regime (orange full circle) present in both Fig. 1a and Fig. 1b, $T_d$ and $R_t$ fluctuate randomly around constant values for a finite time period. Indeed, in contrast with the 1st and 2nd wave, obeying to the analytical law (2) in the supercritical region, in the metastable phase an incoherent disordered behavior emerges. This is evident in the ($T_d$, $R_t$) and ($T_d$, $R_t$-1) projections in Fig. 1c and Fig. 1d, respectively. In panel (d) the dashed line represents the power law best fit of the supercritical regime dataset. The behavior is described in Fig. 1e by showing the 3D plot ($T_d$, $R_t$, t) where the doubling time $T_d$(t, $R_t$) is plotted vs. days and the effective reproductive number $R_t$ for Italy (blue) and Germany (red). Here the metastable phases are followed by the falling in the supercritical regime that marks the raise of the 2nd pandemic wave (dashed arrow). The same mechanism is described by the ($T_d$, $R_t$-1, t) phase diagram of Fig. 1f.

*2.2 Metastable phases in the plateaus before and after the second wave in Italy and Italian regions*

The Covid-19 spreading rate with containment measures (CSRwCM) in a heterogeneous population can be understood in more detail looking at the analysis and the comparison of the pandemic spreading rate in the different Italian regions. Figure 2 shows the 3D ($T_d$, $R_t$, t) phase diagram for three sets of Italian regions where different metastable phases can be recognized. We have combined Emilia Romagna, Toscana and Lazio (left panels), Lombardia, Veneto, and Liguria (central panels), Campania, Puglia and Sicilia (right panels). The semi-logarithmic projection ($T_d$, t) for these groups of regions are shown in Fig 2a, Fig 2b and Fig 2c. Again, in the supercritical regime, 1st and 2nd waves obey the exponential growth of the Eq. 1 as indicated by dashed arrows. Since the growth rate is described by the *s*-factor, we point out that in the 1st wave the exponential epidemics growth is the same for all regions. On the contrary, the 2nd wave exhibit different growth rate, as indicated by the dashed arrows, and are characterized by different s-factors and different slopes. This is due by the different containment measures enforced by the local governments.

The orange areas identify the disordered metastable phases above the critical regime (grey rectangle) with the precursor states of the 2nd phase. Fig 2d, Fig 2e and **Fig 2f** are the ($T_d$, $R_t$) projections confirming the universality of the scaling law in Eq. 2 in the supercritical phase, in all regions. Also in this case, above the critical phase we identify arrested and metastable disordered phases. Finally, Fig 2g, Fig 2h and Fig 2i show the ($T_d$, $R_t$, t) 3D plot for the three sets of Italian regions and the different metastable phases, all ending down the raising point of the 2$^{nd}$ wave.

The time dependence of $T_d$ and $R_t$ values in nine selected Italian regions is compared in Figure 3. In the supercritical regime (yellow areas) we can distinguish both the first and the second pandemic wave where the doubling time (blue curve) increases in the range $2<T_d<40$ days and the reproductive number $R_t$ decreases down to 1. We omitted data of arrested phases, where $T_d$ and $R_t$ fluctuate around their maximum and minimum values, respectively. The green area identifies the metastable phases from where the epidemic spreading can evolve towards a new epidemic wave or towards an arrested phase if an effective containment policy is enforced.

*2.3 The RTD factor to monitor the pandemic dynamics*

In order to take into account both $R_t$ and $T_d$ to describe the pandemic complex dynamics, we introduce for the first time the new parameter

$$RTD = R_t(100/T_d) \qquad (3)$$

Figure 4 shows the RTD parameter as a function of time for nine selected Italian regions. Using this parameter, the pandemic waves occur when RTD>1 (yellow areas). The green area with RTD<1 represents the arrested phase where $R_t<1$ and $T_d>100$. We can also distinguish the metastable phases: the first is precursor of the second pandemic wave while the second is in full swing. In these phases, RTD fluctuates around 1 for a finite time. In the different phases the number of active cases increases during the pandemic waves and decreases in the arrested phase. In the metastable phases RTD exhibits a slightly flat behavior around the value of one.

In Figure 5 we compare maps of the Italian regions and the RTD values for different days selected in different phases. When RTD >1 the system enters the critical regime. The color evolution from yellow to red corresponds to $T_d$ and $R_t$ values while spreading the pandemic. In particular, we can visualize when in each region RTD approaches the critical regime (red ~~colour~~), and thus when a prompt intervention of policymakers and institutions is mandatory. Indeed, any delayed decision may affect directly both the duration of the pandemic and the number of fatalities.

## 3. Conclusions

We have provided here an original quantitative approach to describe and understand the time evolution of the Covid-19 pandemic, characterizing in a quantitative way the evolution stages. We show that it is necessary to expand the parameter space, combining relevant variables like $T_d$, $R_t$ in the new parameter RTD = $R_t(100/T_d)$, to precisely monitor the evolution of the pandemic. This approach makes possible to analyse the dynamics, probing and tuning at the same time containment measures. RTD sheds light and provides a new quantitative experimental tool not only for the quantitative statistical physics of the Covid-19 pandemic, but thanks to its predicting power also for any future epidemic events. Our results, showing the presence of a metastable plateaus at the end of the second wave, support the recent evidence[14] that susceptible population and dynamic heterogeneity over multiple timescales of individual variations in social activity induces in the epidemic dynamics the succession of epidemics waves separated by metastable states of transient collective immunity (TCI), which is a fragile state that abruptly turns to the supercritical regime of the third wave following short timescales super-spreading events, associated with loosing containment rules in the LSG policy.

The Italian policy of Covid-19 containment measures in 2021 has followed the Imperial College protocol called Lockdown Stop and Go (LSG) with keeping the period "Go" as long as the number of available beds in Intensive care units of hospitals was below saturation, followed by "Stop" periods. On the contrary, in many other countries, e.g., in the pacific-asian region, the L*ockdown case Finding mobile Tracing* (LFT) policy has been enforced using a strict Quarantine, characterized by mandatory mobile contact tracing with rollout of mass testing. The policy was to reduce the number of infections below the tens per million populations with a different target called "*Zero infections*". The comparative evaluation of the LSG versus LFT protocols can be summarized by the positive economic trend observed in LFT countries (e.g. South Korea or China)[23,29-33] and the negative economic trend in LSG countries like Italy. The economic losses have been due to the long stop of manufacture activities in LSG countries during the supercritical regime periods. In these countries lockdown has been applied without mandatory contact tracing and loose quarantine during the first and second wave. On the contrary in LFT countries the supercritical regime time periods have been a factor three time shorter with a much less impact on the PIL.

The economic losses follow the number of fatalities per million populations, which are more than a factor 100 in LSG versus LFT countries[23,29-33]. The current high number of daily fatalities ~~per day~~ in the metastable regime near criticality and the shortage of vaccines require the urgent adoption of a new plan taking into account the evolution of the pandemic, which occurs above all in the active population. It is therefore mandatory to act looking at the "Zero infections" target by enforcing strict contact tracing[23,29-33] and quarantine rules, by stopping the emergence of new virus variants in the near zero infections regime during the long vaccination campaign of the entire population. This new policy will ~~to~~ speed up economy like in the Pacific Asian area and will reduce the still large number of daily fatalities per million populations. The rapid detection of any possible onset of the third wave by looking at the RTD parameter evolution will allow the prompt political response avoiding the new lockdowns predicted in the LSG strategy. The use of the RTD parameter in the containment policy will provide early warning to critical situations slowing down the rate of diffusion of the virus during the vaccination campaigns, in particular for the scarce resources of vaccines and the appearance of new virus variants, saving economic losses and reducing the number of fatalities.

**Acknowledgments**

This work was funded by Superstripes onlus. M V Mazziotti acknowledge the financial support of *Superstripes onlus*.

**Figures**

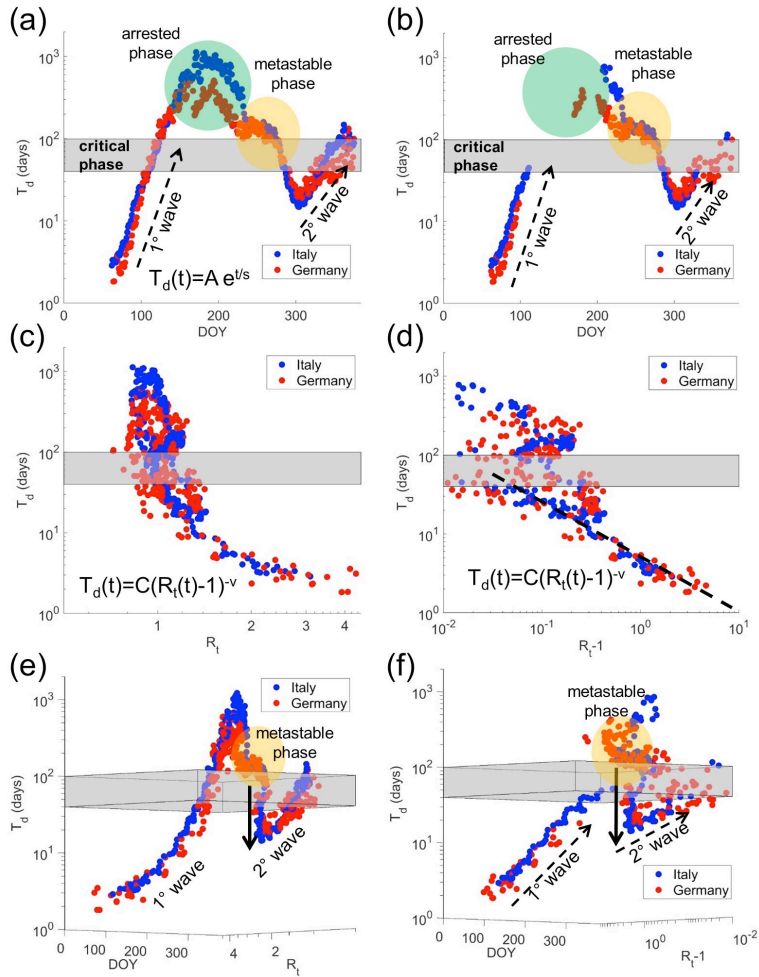

**Figure 1 - 3D Phase diagram ($T_d$, $R_t$, t) for Italy (blue) and Germany (red)**
**(a)** Semi-logarithmic ($T_d$, t) projection of the 3D Phase Diagram ($T_d$, $R_t$, t). Is evident the supercritical exponential growth $T_d(t)=Ae^{t/s}$ [dashed arrows] in the 1$^{st}$ and 2$^{nd}$ wave. The exponential growth is arrested above the critical phase (grey area) where $T_d$ >100 and $R_t$ becomes < 1 [green circles]. This is also seen in the semi-logarithmic projection ($T_d$, $R_t$-1) in panel **(b)** where values $R_t$<1 are cancelled. After the arrested phase, if $T_d$ growths and $R_t$ increases a second pandemic phase occurs. The precursors of the new phase are observed in the region with $T_d$>100, over the critical phase, and $R_t$ >1. This area is the orange circle in panel (a) and (b). In the supercritical region both the 1$^{st}$ and 2$^{nd}$ wave obey to the same mathematical law [$T_d(t)=C(R_t(t)-1)^{-\nu}$] as shown in the ($T_d$, $R_t$) and ($T_d$, $R_t$-1) projections in panel **(c)** and **(d)**, respectively. In panel **(d)** the dashed line represents the power law best fit of data in the supercritical regime. The critical phase [40<$T_d$<100 days] is outlined by the horizontal grey dashed strip. Above the critical regime [$T_d$>100 days and $R_t$<1] an incoherent disordered behavior is observed. **(e)** The ($T_d$, $R_t$, t) 3D phase diagram for Italy (blue) and Germany (red) where the doubling time $T_d(t, R_t)$ is plotted vs. DOY and the effective reproductive number $R_t$. Also here the gray space outlines the critical crossover [40<$T_d$<100 days] and separates the supercritical phase [$T_d$<40 days; $R_t$>1] from the arrested subcritical phase [$T_d$>100 days; $R_t$<1]. We indicate here the metastable phase followed by the fall in the supercritical regime and the raising of the 2$^{nd}$ pandemic wave (dashed arrow). The same behavior is described in the ($T_d$, $R_t$-1, t) phase diagram in panel **(f)**.

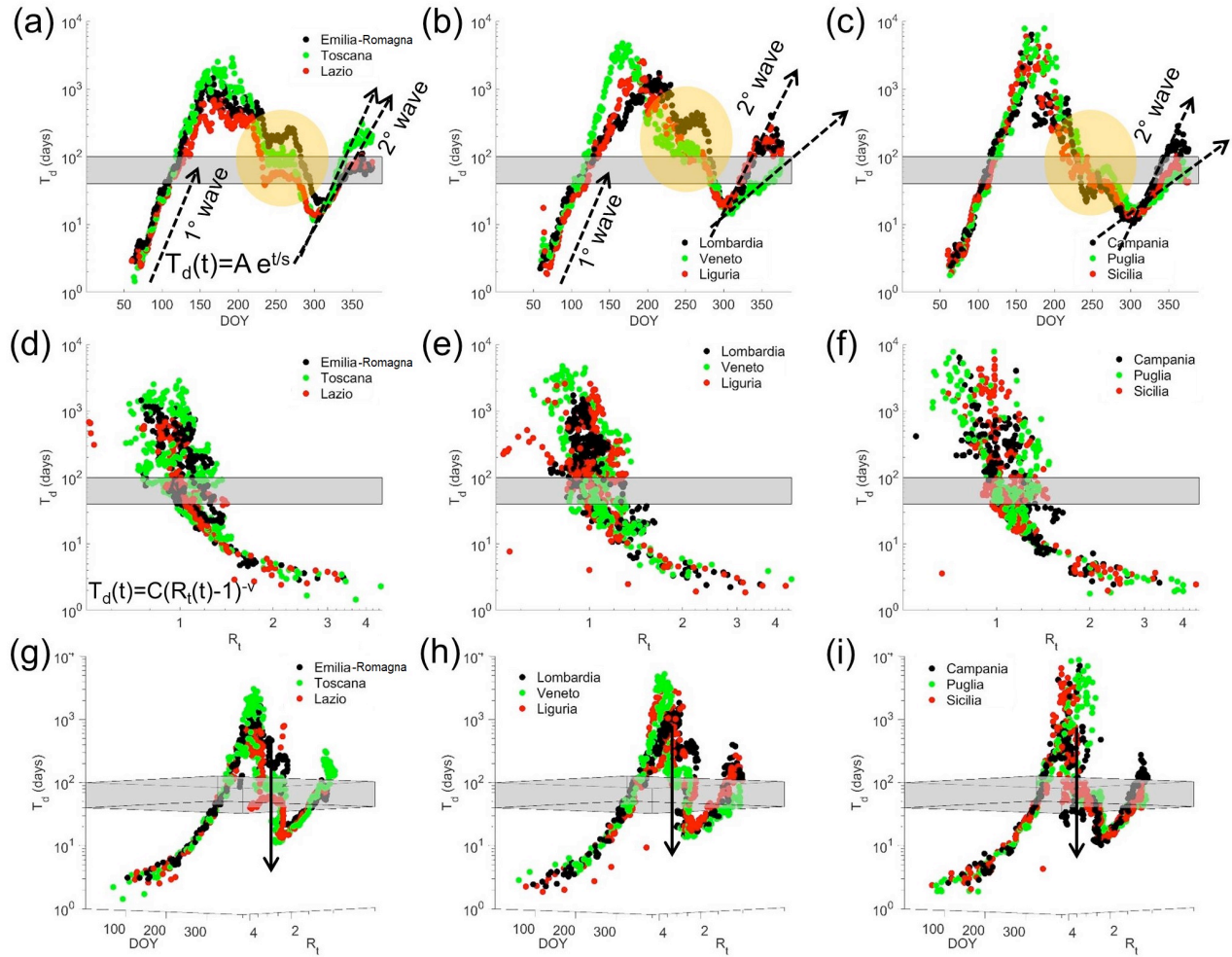

**Figure 2.**
The ($T_d$, $R_t$, t) 3D phase diagram for several Italian regions. The planar semi-logarithmic projection ($T_d$, t) for three different groups of regions are shown in panel **(a)**, **(b)** and **(c)**. The 1st and 2nd waves obeying the exponential growth [$T_d(t)=Ae^{t/s}$] are outlined by dashed arrows in the supercritical regime. Since the growth rate is described by the *s*-factor, in the 1st wave the exponential epidemics growth is the same for all regions. On the contrary, the 2nd wave exhibits different growth rate due to the different *s*-factor [dashed arrows], with different slopes. The orange areas outline the disordered metastable phases above the critical regime (grey rectangle) and the precursors of the 2nd phase. The panels **(d)**, **(e)** and **(f)** are the ($T_d$, $R_t$) projections showing the universality of the scaling law $T_d(t)=C(R_t(t)-1)^{-v}$ in the supercritical phase in all regions. Also in this case we observe the arrested and metastable disordered phases above the critical phase. Finally, in panels **(g)**, **(h)** and **(i)** are showed the ($T_d$, $R_t$, t) 3D phase diagram of three groups of Italian regions to recognize the different metastable phases all ending around the raising point of the 2nd wave.

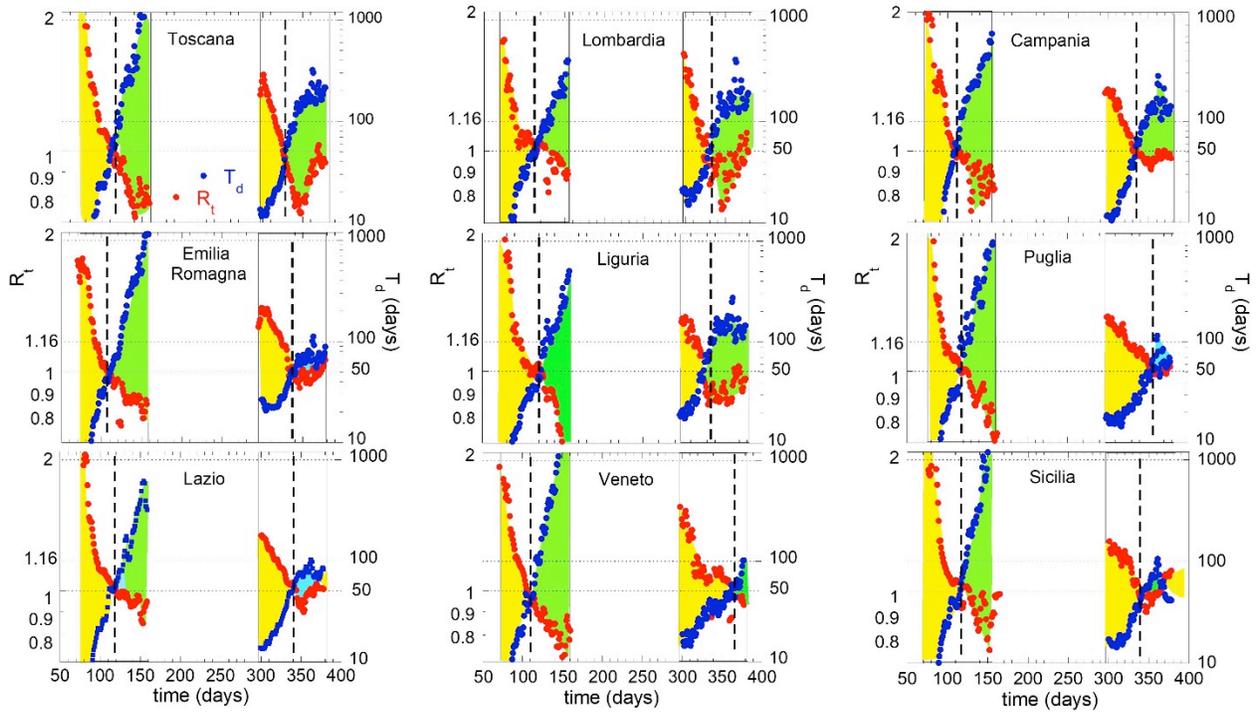

**Figure 3**

$T_d$ and $R_t$ vs. time in nine selected Italian regions. In the supercritical regime [yellow areas] the doubling time (red) increases in the range $2<T_d<40$ days while the reproductive number $R_t$ decreases to 1. We omitted data of the arrested phases, where $T_d$ and $R_t$ fluctuate around maximum and minimum values, respectively. The green area identifies the metastable phases from where the epidemic spreading may evolve towards an arrested phase or a new epidemic wave.

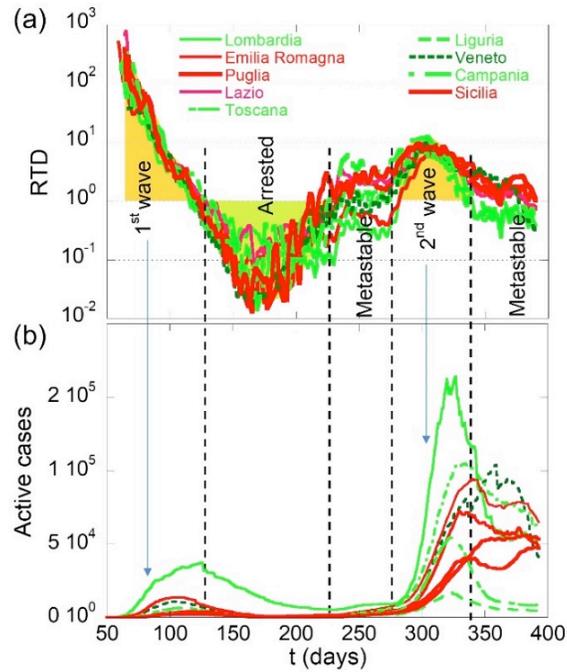

**Figure 4**.
(a) RTD vs. time for the Italian regions. The two pandemic waves are described by the two yellow areas among dashed vertical lines where RTD>1. The green area [RTD<1] outlines the arrested phase where $R_t<1$ and $T_d>100$. We can distinguish the first metastable phases and the precursors of the second pandemic wave while the second metastable phase is in full swing. In these phases, RTD fluctuates around 1 for a finite time. (b) The cumulative curve of the active cases per millions of populations. We observe the increase during the pandemic waves as well as the decrease in the arrested phase and the almost flat behavior in the metastable phases.

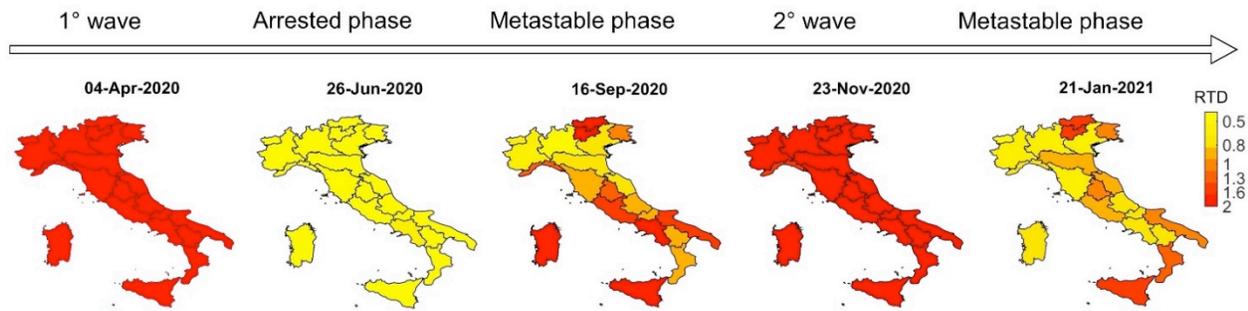

**Figure 5.**
Geographic maps of Italian regions at different time in different phases with their local RTD values. The color bar evolution from yellow to red is due to $T_d$ and $R_t$ values and to the spreading of pandemics.